\documentclass[aps,prx,reprint,superscriptaddress]{revtex4-2}

\usepackage[T1]{fontenc}
\usepackage{amsmath}
\usepackage{graphicx}
\usepackage{xcolor}
\usepackage[noend]{algpseudocode}
\usepackage{algorithm}
\usepackage{qcircuit}
\usepackage{hyperref}
\usepackage[normalem]{ulem}
\usepackage{dsfont}

\makeatletter
\def\BState{\State\hskip-\ALG@thistlm}
\makeatother

\def\ket#1{\left| #1 \right\rangle}

\begin{document}

%\preprint{}

\title{Preparing Bethe Ansatz Eigenstates on a Quantum Computer}


\author{John S. Van Dyke}
%\thanks{}
\affiliation{Department of Physics, Virginia Tech, Blacksburg, VA 24061}

\author{George S. Barron}
\affiliation{Department of Physics, Virginia Tech, Blacksburg, VA 24061}

\author{Nicholas J. Mayhall}
\affiliation{Department of Chemistry, Virginia Tech, Blacksburg, VA 24061}

\author{Edwin Barnes}
\affiliation{Department of Physics, Virginia Tech, Blacksburg, VA 24061}

\author{Sophia E. Economou}
\affiliation{Department of Physics, Virginia Tech, Blacksburg, VA 24061}


\date{\today}

\begin{abstract}
Several quantum many-body models in one dimension possess exact solutions via the Bethe ansatz method, which has been highly successful for understanding their behavior.  Nevertheless, there remain physical properties of such models for which analytic results are unavailable, and which are also not well-described by approximate numerical methods. Preparing Bethe ansatz eigenstates directly on a quantum computer would allow straightforward extraction of these quantities via measurement. We present a quantum algorithm for preparing Bethe ansatz eigenstates of the spin-1/2 XXZ spin chain that correspond to real-valued solutions of the Bethe equations.  The algorithm is polynomial in the number of T gates and circuit depth, with modest constant prefactors. Although the algorithm is probabilistic, with a success rate that decreases with increasing eigenstate energy, we employ amplitude amplification to boost the success probability. The resource requirements for our approach are lower than other state-of-the-art quantum simulation algorithms for small error-corrected devices, and thus may offer an alternative and computationally less-demanding demonstration of quantum advantage for physically relevant problems.
\end{abstract}

%\keywords{}

\maketitle


%\subsection{}
%\subsubsection{}

Quantum computers hold the promise of transformative applications in a variety of fields including cryptanalysis \cite{Shor1994}, quantum chemistry \cite{Aspuru-Guzik2005,McArdle2020}, materials science \cite{Abrams1997,Wecker2015b}, and potentially combinatorial optimization \cite{Grover1997,Farhi2014}. To realize the full potential of quantum computing, large-scale, fault-tolerant devices will ultimately be necessary. As these do not yet exist, much recent work has studied possible near-term applications in the present era of noisy, intermediate-scale quantum computers (NISQ) \cite{Preskill2018,Deutsch2020}. In this context, a key question concerns the demonstration of `quantum advantage' -- that is, the ability to perform computations that can not be done efficienly with classical methods. Recently, quantum advantage was shown for a superconducting processor sampling random quantum circuits \cite{Arute2019compact} and photonic-based Gaussian boson sampling \cite{Zhong2020}.  Although these are important achievements, the specific tasks performed were not closely related to the practical applications mentioned above, but were essentially designed for the purpose of demonstrating advantage.  Thus, the realization of quantum advantage for a problem of practical interest remains open.

A question of increasing interest is what applications become feasible with small-scale error-corrected devices, i.e., in the intermediate era between NISQ and fault-tolerant quantum computers with many logical qubits and a high clock speed for non-Clifford gates. Recent estimates suggest that algorithms that provide only a quadratic speedup over classical methods may have difficulty achieving quantum advantage on problem sizes accessible with small devices \cite{Campbell2019,Sanders2020,Babbush2020}. The simulation of quantum systems, on the other hand, can yield exponential improvement over conventional approaches. These still require formidable resources, despite recent algorithmic advances \cite{Babbush2018,Kivlichan2020,vonBurg2020,Lee2020}. This motivates the search for physically-interesting problems and algorithms that can lead to quantum advantage with fewer resources in the near future.  

We propose the study of Bethe ansatz (BA) states on a quantum computer as a computationally less-demanding route to the demonstration of quantum advantage for problems relevant to physics, including quantum magnetism \cite{Nagler1991,Wang2018}, ultracold atoms \cite{Fukuhara2013}, and unconventional superconductivity \cite{Pasnoori2020}. BA methods allow for deep insight into the static and dynamic properties of these many-body systems, and are able to explore not only ground states, but also interactions between complex collective excitations, such as magnons, and the response to quantum quench experiments \cite{Calabrese2006}.

More specifically, the BA technique yields exact solutions to a class of one-dimensional quantum many-body models, including the spin-1/2 Heisenberg and Hubbard models, among others \cite{Bethe1931,Lieb1968,Mattis1993,Essler2005}.  The resulting wave functions depend on algebraic equations that can be efficiently solved classically. The exponential growth of the Hilbert space with the system size $L$ has historically limited the direct computational studies of the eigenstates to small systems.  Instead, various mathematical techniques have been extensively developed to access physical quantities in the thermodynamic limit $L\rightarrow\infty$, bypassing the calculation of the wave function itself.  While many different quantities can be determined, the difficulty of their calculation varies widely.  In particular, arbitrary-range and higher-order correlation functions have been very challenging to access, and remain an active area of research \cite{Gohmann2017,Babenko2021,Babenko2020}.  Quantum computers, however, can compute such correlation functions \cite{Somma2002,Wecker2015b} straightforwardly, thus suggesting the possibility of quantum advantage for this task.  The importance of higher-order correlation functions for strongly correlated systems has recently been emphasized \cite{Bohrdt2021}. Calculating such observables using a quantum computer in turn hinges on the possibility of efficiently preparing the Bethe ansatz states. 

To this end, we demonstrate an efficient quantum algorithm that can prepare a subset of the Bethe ansatz eigenstates of the one-dimensional XXZ chain, a model which is fundamental to the study of quantum magnetism. Our algorithm uses the so-called coordinate Bethe ansatz method, and has polynomial scaling in the circuit depth and T-gate complexity, along with low constant prefactors.  As we show with explicit gate counts for the corresponding circuits, the approach scales to large enough systems for the calculation of classically inaccessible quantities in near-term error-corrected devices.  While the algorithm we present is probabilistic, we also show that amplitude amplification can be used to increase the success rate \cite{Brassard2000}.

Algorithms have been previously given (and also implemented) that prepare exact eigenstates of quantum many-body models \cite{Verstraete2009,Schmoll2017,Cervera-Lierta2018,Robbins2021}. However, these were largely limited to cases that are equivalent to non-interacting fermions, for instance, under the Jordan-Wigner mapping. In contrast, the XXZ model corresponds to an interacting fermionic problem, which is computationally much more difficult.  We note that the possibility of constructing circuits to diagonalize Bethe ansatz-solvable models and measure challenging correlation functions was previously suggested, though without an indication of how this could be done \cite{Verstraete2009}.

The importance of Bethe ansatz-solvable models for benchmarking NISQ devices has been previously recognized \cite{Dallaire-Demers2020,Cervia2020,Robbins2021}, as they provide exact values for quantities (such as the energy) to compare against the results of noisy quantum computations.  On the other hand, the direct preparation of Bethe ansatz states has been relatively unexplored.  This question was recently studied in Ref. \cite{Nepomechie2021}, which considered treating Bethe ansatz states variationally (using the algebraic Bethe ansatz) and concluded that the approach was not scalable.  Furthermore, that work did not make a connection to the possibility of quantum advantage.  Apart from direct preparation of Bethe ansatz eigenstates, other works have considered variational approaches using generic ansatzes \cite{Dallaire-Demers2020,Cervia2020,Ho2019,Wiersema2020,Cervera-Lierta2020}. The comparison of the computational complexity of these methods to that of the direct construction is an interesting question for future studies, as are probabilistic algorithms for preparing other strongly-correlated states \cite{Murta2021}.

The paper is organized as follows.  Section \ref{sec:model} introduces the XXZ model and the elements of the Bethe ansatz solution needed for the construction of the algorithm.  Section \ref{sec:bethealg} describes the Bethe ansatz state preparation algorithm.  Section \ref{sec:numerics} presents numerical results that validate the method and studies its success probability.  This section also includes resource estimates for classically intractable problem sizes.  Section \ref{sec:amplitudeamp} describes the amplitude amplification procedure for our algorithm, and presents numerical calculations that  confirm its success.  Section \ref{sec:comparealg} compares our algorithm with conceptually simpler but less efficient approaches to the same task, explicitly verifying the enormous speedup of our method.  Section \ref{sec:discussion} argues that quantum advantage can be achieved with Bethe state preparation by comparison with classical computational methods, and presents additional applications of the algorithm.  Finally, we conclude in Section \ref{sec:conclusions}.

\section{Model and Solution \label{sec:model}}

We consider the one-dimensional spin-1/2 XXZ chain on $L$ sites with periodic boundary conditions, whose Hamiltonian is given by
\begin{align}
H =  \sum_{i=0}^{L-1} J_{xy} \left( S^x_i S^x_{i+1} + S^y_i S^y_{i+1} \right) + J_z S^z_i S^z_{i+1},
\end{align}
with $S^\alpha_L \equiv S^\alpha_0$ ($\alpha = x,y,z$).  Here $S^\alpha_j$ are the spin operators with eigenvalues $\pm 1/2$. The exact solution of this model via the Bethe ansatz method was presented in Ref. \cite{Orbach1958}, and many introductions to the problem (in both the coordinate and algebraic formulations) exist \cite{Korepin1993,Takahashi1999,Giamarchi2004,Sutherland2004,Gaudin2014}. We follow the account of the Bethe ansatz method given in Ref. \cite{Giamarchi2004}.  The eigenstates of the above Hamiltonian are given by 
\begin{align}
\psi(x_1,\dots,x_M) = \sum_P A_P \exp \left[ i \sum_{j=1}^M k_{Pj} x_j \right], \label{eq:bethewavefunction}
\end{align}
where $x_1,\dots,x_M$ label the positions of the $M$ down spins in the chain (the Hamiltonian conserves the $z$ component of the total spin, $S^z_{tot} = \sum_i S^z_i$) and the momenta $k_i$ label the different states.  The wave function of Eq. \eqref{eq:bethewavefunction} gives the amplitude for the $M$ down spins to occur on the sites $x_j = 0,\dots,L\!-\!1$.  The summation here is over the $M!$ permutations of the down spin sites.  These permutations arise from the fact that, within Bethe ansatz models, scattering processes exchange momenta between particles, but do not alter their magnitudes.  The coefficients $A_P$ are related by
\begin{align}
\frac{A_P}{A_{P'}} &= - \frac{1 + e^{i (k_{Pl} + k_{P'l})} - 2 \Delta e^{ik_{Pl}}} {1 + e^{i (k_{Pl} + k_{P'l})} - 2 \Delta e^{ik_{P'l}}} \notag\\
&\equiv - e^{-i \Theta (k_{Pl},k_{P'l})}. \label{eq:AP}
\end{align}
To fix the coefficients, we take $A_{I} = 1$, where $I$ is the identity permutation.  Here $\Delta = J_z/J_{xy}$ is the anisotropy in the interactions, and $P$ and $P'$ are permutations that differ by a single transposition between adjacent elements, $P(l+1)=P'(l)$ and $P(l)=P'(l+1)$.  The momenta $k_i$ are constrained by the quantization conditions,
\begin{align}
L k_i = 2\pi I_i + \sum_j \Theta (k_i,k_j), \label{eq:quantization}
\end{align}
with $I_i$ an integer (half-integer) for $M$ odd (even).  Physically, these constraints on $k_i$ arise from imposing periodic boundary conditions on the model.  Eqs. \eqref{eq:AP} and \eqref{eq:quantization} are a set of algebraic equations (the Bethe equations) for the quantum numbers $\{k_i\}$.  In general, these equations admit complex solutions, but for the special case when all $\{k_i\}$ are real, $\Theta(k_i,k_j)$ is also real and is given by
\begin{align}
\Theta (k_i,k_j) = 2 \arctan \left( \frac{\Delta \sin (\frac{k_i-k_j}{2})}{\Delta \cos (\frac{k_i-k_j}{2}) - \cos (\frac{k_i+k_j}{2})} \right).
\end{align}

In the spirit of hybrid quantum-classical algorithms, we solve the Bethe equations classically to obtain the momenta $\{k_i\}$ and phases $\Theta(k_i,k_j)$. These values are then used as input to our quantum algorithm for generating the corresponding eigenstate. Our algorithm allows for the preparation of Bethe ansatz eigenstates for which the $\{k_i\}$ are real, such that $A_P$ and $e^{i k_{Pj} x_j}$ amount to complex phases applied to the second-quantized basis states of the system.

\section{Bethe Ansatz State Preparation Algorithm \label{sec:bethealg}}

The quantum algorithm for preparing a Bethe ansatz state consists of several steps, and is summarized in Algorithm \ref{alg:bethe}. The general structure of our approach is based on the linear combination of unitaries (LCU) method (which also finds application in the Taylor series approach to Hamiltonian simulation) \cite{Childs2012,Berry2015}. However, a key difference is that our algorithm aims to generate specific quantum states starting from a particular initial state, rather than compiling a generic unitary evolution operator. In addition to the $L$ qubits representing the system, a register of $M^2$ ancilla qubits are used to label the different permutation terms in Eq.~\eqref{eq:bethewavefunction}. By preparing a superposition of the allowed label values on these ancillas, using these to apply controlled operations on the system, and finally disentangling the label and system registers, we perform the summation over all permutations present in Eq.~\eqref{eq:bethewavefunction}. This process is facilitated by introducing a second ancillary register of $M$ qubits that we call the ``faucet register", along with one additional ancilla work qubit. Thus, the algorithm requires a total of $M^2+M+1$ ancilla qubits. 

\begin{algorithm}[H]
\caption{Bethe state preparation}\label{alg:bethe}
\begin{algorithmic}[1]
\State Prepare the Dicke state $|D_{L,M}\rangle$ on the system qubits
\State Create permutation labels while applying pieces of $A_P$
\State Apply $e^{ik_{Pj}x_j}$ using the ``faucet'' method
\State Reverse permutation label (without phases)
\State Measure permutation label, with success on $|00\cdots0\rangle$
\end{algorithmic}
\end{algorithm}

The algorithm begins by preparing the Dicke state on $L$ sites with $M$ down spins, $|D_{L,M}\rangle$.  Relabeling $|\!\!\uparrow\rangle \equiv |0\rangle$, $|\!\!\downarrow\rangle \equiv |1\rangle$, $|D_{L,M}\rangle$ is the equal superposition (that is, without relative phases) of all basis states on $L$ qubits with Hamming weight $M$.  This state forms the underlying ``canvas'' on which the phases in Eq.~\eqref{eq:bethewavefunction} are applied.  Dicke state preparation can be accomplished using the recent deterministic algorithm of Ref. \cite{Bartschi2019}, for which the gate count was improved in Ref. \cite{Mukherjee2020}.  This algorithm uses an inductive method to prepare smaller Dicke states which are subsequently combined to yield the desired one.  We have used this algorithm in our explicit circuit constructions, though any other deterministic method of preparing $|D_{L,M}\rangle$ would also work.  

As discussed above, the amplitudes that must be applied to $|D_{L,M}\rangle$ to generate a Bethe ansatz state depend on the permutations $\{P\}$ of $M$ objects.  We use the permutation label register to create the different permutations and their associated phases $A_P$. Naively, one could use an integer encoded in a binary representation to label each of the permutations.  The difficulty with this approach is that the number of permutations is $M!$, so that imprinting the phases $A_P$ and $e^{i k_{Pj} x_j}$ onto $|D_{L,M}\rangle$ would require combinatorially many operations.  This leads to circuit depths and complexities that are superexponential in $M$, quickly becoming unfeasible as $M$ grows (we explore a concrete realization of this approach in Section \ref{sec:comparealg}).  To overcome this fundamental limitation of this method and design an efficient algorithm, we introduce a conceptually distinct approach for labeling the permutations.  Rather than assigning an arbitrary number to a given permutation, we implement its explicit action on the string consisting of the numbers $1,\dots,M$.  As described below, this allows for an efficient generation of the permutation labels, while also generating the distinct $A_P$ simultaneously.  

The permutation label register consists of $M$ subregisters, each of which can store an integer value $k\in\{1,\dots,M\}$.  To represent a valid permutation, the subregisters must contain distinct values (for instance, $|213\rangle$ is valid whereas $|233\rangle$ is not).  We use a one-hot encoding such that each subregister consists of $M$ qubits, and the number $k$ is represented by a $1$ on the $k$th qubit and $0$s on the rest.  Thus, for $M=3$ the allowed states on each subregister are $|1\rangle \equiv |001\rangle$, $|2\rangle \equiv |010\rangle$, and $|3\rangle \equiv |100\rangle$.  This one-hot encoding requires $M^2$ qubits to represent the complete label.  The use of the one-hot encoding facilitates a trade-off between time and space resources \cite{Childs2018,Babbush2018,Wan2021} by reducing the number of controls required to implement the necessary phase gates.

\begin{figure*}[]
\includegraphics[scale=0.52]{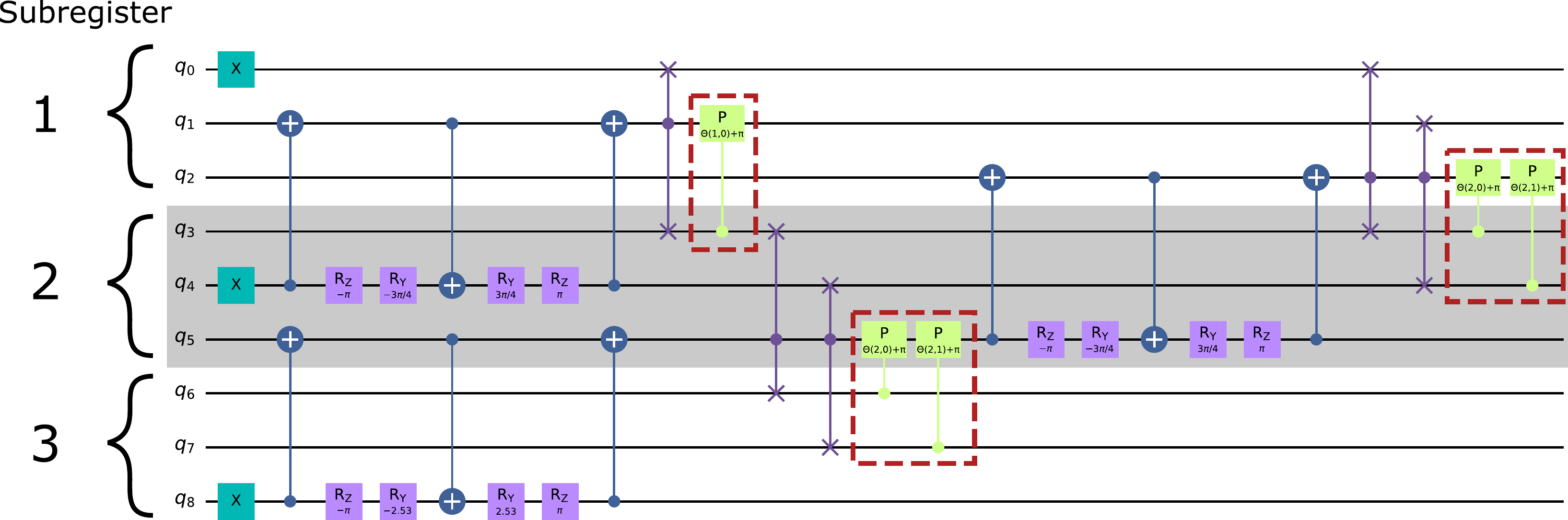}
\caption{Circuit to prepare the permutation label ancilla state for $M=3$.  When the gates inside the red dashed rectangles are excluded, the circuit realizes the equally-weighted superposition encoding the permutations of three objects. With these gates included, the circuit applies the additional permutation-dependent phases $A_P$. The three subregisters are indicated by shading, and the initial state of each qubit is $\ket{0}$. The alternating \textsc{cnot} gates interleaved with $R_y$ and $R_z$ rotations realize the \textsc{aswap} gates.  \label{fig:permlabelM3circuit}} 
\end{figure*}

The goal of step 2 of Algorithm~\ref{alg:bethe} is to create the state $\frac{1}{\sqrt{M!}} \sum_P A_P \ket{P}$ on the permutation label register. The phases $A_P$ are kicked back onto the system qubits, while the $|P\rangle$ are used to apply the conditional gates needed in step 3, as explained below. For clarity, we first describe the construction of the equal superposition of all permutation labels.  We then show how to slightly modify this procedure to simultaneously generate the phases $A_P$ for all $M!$ permutations.  We use an iterative method to construct the permutation label state starting from the vacuum state $|00\dots0\rangle$ on $M^2$ qubits.  The complete label superposition state is built up sequentially from the first (rightmost) subregister to the last (leftmost) using a series of exchange-type gates.  We describe the method inductively as follows.  The zeroth sublabel is prepared by setting the zeroth qubit of the zeroth subregister to 1 (in the following, the index $k$ is enumerated starting from 0).  Assume the $k$th sublabel (i.e., an equal superposition of permutations of integers 1 through $k+1$) has been constructed on the $k+1$ rightmost subregisters.  Set the $(k+1)$th qubit of the $(k+1)$th subregister to 1, thus introducing the next integer value to be included in the permutation label state.  Perform the exchange-type \textsc{aswap} gate \cite{Barkoutsos2018,Gard2020}, 
\begin{align}
A(\theta,\phi) = \begin{pmatrix}
1 & 0 & 0 & 0\\
0 & \cos (\theta) & e^{i \phi} \sin (\theta) & 0\\
0 & e^{-i \phi} \sin (\theta) & -\cos( \theta) & 0\\
0 & 0 & 0 & 1
\end{pmatrix},
\end{align}
between the $(k+1)$th qubits of subregisters $k+1$ and $k$, with $\theta=\arccos (1/\sqrt{k+2})$, $\phi=0$. This generates a superposition state consisting of two sets of terms: those in which the 1 remains in the $(k+1)$th subregister and those in which it is transferred to the $k$th.  In the latter case, the $(k+1)$th subregister now contains all 0s, while the $k$th has two qubits with 1, which is not valid.  This is fixed by applying controlled-\textsc{swap} gates between all qubits $l < k+1$ in subregisters $k$ and $k+1$, controlled on the state of qubit $k+1$ in subregister $k$.  Taken together, these operations produce a partial \textsc{swap} between subregisters $k+1$ and $k$,
 \begin{align}
&\ket{k\!+\!1}_{k+1}\ket{m}_{k} \notag \\
&\to \frac{1}{\sqrt{k\!+\!2}}\ket{k\!+\!1}_{k+1}\ket{m}_{k}+\sqrt{\frac{k\!+\!1}{k\!+\!2}}\ket{m}_{k+1}\ket{k\!+\!1}_{k}, \label{eq:swapstate}
 \end{align}
 where $\ket{i}_j$ is the one-hot encoded state for $i$ on subregister $j$, and $m < k$ by construction. One repeats this partial swapping process, now between subregisters $k$ and $k-1$, then between $k-1$ and $k-2$, and so on, until the last register has been swapped.  By implementing the inductive process up to the $(M-1)$th subregister, the complete equally-weighted superposition of permutation labels is formed.  As an example, for $M=3$ the above algorithm generates the following sequence of state transformations:
% \begin{align}
% &\ket{0}\!\ket{0}\!\ket{0} \to \ket{0}\!\ket{0}\!\ket{1} \to \ket{0}\!\ket{2}\!\ket{1} \notag \\
% &\to \frac{1}{\sqrt{2}}\ket{0} \big( \! \ket{2}\!\ket{1} + \ket{1}\!\ket{2} \!\big) \to \frac{1}{\sqrt{2}} \ket{3} \big(\! \ket{2}\!\ket{1} + \ket{1}\!\ket{2} \!\big) \notag \\
% &\to \frac{1}{\sqrt{6}} \ket{3}\!\ket{2}\!\ket{1} + \frac{1}{\sqrt{3}} \ket{2}\!\ket{3}\!\ket{1} \notag \\
% & \phantom{\to} + \frac{1}{\sqrt{6}} \ket{3}\!\ket{1}\!\ket{2} + \frac{1}{\sqrt{3}} \ket{1}\!\ket{3}\!\ket{2} \notag \\
% &\to \frac{1}{\sqrt{6}}\big(\! \ket{3}\!\ket{2}\!\ket{1} + \ket{2}\!\ket{3}\!\ket{1} + \ket{2}\!\ket{1}\!\ket{3} \notag \\
% &\phantom{\to \frac{1}{\sqrt{6}}\big(\!} + \ket{3}\!\ket{1}\!\ket{2} + \ket{1}\!\ket{3}\!\ket{2} + \ket{1}\!\ket{2}\!\ket{3} \!\big).
% \end{align}
\begin{widetext}
\begin{align}
&\ket{000}\!\ket{000}\!\ket{000} \xrightarrow{\text{X}} \ket{000}\!\ket{000}\!\ket{001} \xrightarrow{\text{X}} \ket{000}\!\ket{010}\!\ket{001} \xrightarrow{\text{ASWAP}} \frac{1}{\sqrt{2}}\ket{000}\big( \ket{010}\!\ket{001} + \ket{000}\!\ket{011}\big)\notag\\
&\xrightarrow{\text{CSWAPs}} \frac{1}{\sqrt{2}}\ket{000}\big(\ket{010}\!\ket{001} + \ket{001}\!\ket{010}\big) \xrightarrow{\text{X}} \frac{1}{\sqrt{2}}\ket{100}\big(\ket{010}\!\ket{001} + \ket{001}\!\ket{010}\big)\notag\\
&\xrightarrow{\text{ASWAP}}\frac{1}{\sqrt{6}}\ket{100}\!\ket{010}\!\ket{001} + \frac{1}{\sqrt{3}}\ket{000}\!\ket{110}\!\ket{001} + \frac{1}{\sqrt{6}}\ket{100}\!\ket{001}\!\ket{010} + \frac{1}{\sqrt{3}}\ket{000}\!\ket{101}\!\ket{010} \notag\\
&\xrightarrow{\text{CSWAPs}}\frac{1}{\sqrt{6}}\ket{100}\!\ket{010}\!\ket{001} + \frac{1}{\sqrt{3}}\ket{010}\!\ket{100}\!\ket{001} + \frac{1}{\sqrt{6}}\ket{100}\!\ket{001}\!\ket{010} + \frac{1}{\sqrt{3}}\ket{001}\!\ket{100}\!\ket{010} \notag\\
&\xrightarrow{\text{ASWAP}} \frac{1}{\sqrt{6}}\ket{100}\!\ket{010}\!\ket{001} + \frac{1}{\sqrt{6}}\ket{010}\!\ket{100}\!\ket{001} + \frac{1}{\sqrt{6}}\ket{010}\!\ket{000}\!\ket{101}+ \frac{1}{\sqrt{6}}\ket{100}\!\ket{001}\!\ket{010} \notag\\
&\phantom{\xrightarrow{\text{ASWAP}}} + \frac{1}{\sqrt{6}}\ket{001}\!\ket{100}\!\ket{010} +
\frac{1}{\sqrt{6}}\ket{001}\!\ket{000}\!\ket{110} \notag\\
&\xrightarrow{\text{CSWAPs}} \frac{1}{\sqrt{6}}\ket{100}\!\ket{010}\!\ket{001} + \frac{1}{\sqrt{6}}\ket{010}\!\ket{100}\!\ket{001} + \frac{1}{\sqrt{6}}\ket{010}\!\ket{001}\!\ket{100}+ \frac{1}{\sqrt{6}}\ket{100}\!\ket{001}\!\ket{010} \notag\\
&\phantom{\xrightarrow{\text{ASWAP}}} + \frac{1}{\sqrt{6}}\ket{001}\!\ket{100}\!\ket{010} +
\frac{1}{\sqrt{6}}\ket{001}\!\ket{010}\!\ket{100}
\end{align}
\end{widetext}
The explicit circuit for $M=3$ is shown in Fig. \ref{fig:permlabelM3circuit}, where the gates inside the red dashed rectangles are excluded at this point.  Furthermore, gates acting on distinct qubits have been pushed to the left, thereby reducing the circuit depth.  This produces a different sequence of intermediate states than above, but the final state is the same.

A slight modification of the above procedure allows one to simultaneously apply the phases $A_P$ to the appropriate terms in the superposition.  After each partial subregister swap, one applies a controlled phase gate with angle $\Theta (k_{i},k_{j}) + \pi$, where $i,j$ are the integer values that have been swapped, and where the additional phase $\pi$ implements the signature of the permutation (red dashed rectangles in Fig.~\ref{fig:permlabelM3circuit}). Suppose, as in Eq. \eqref{eq:swapstate}, that the value $(k+1)$ is swapped to the right. Then the $(k+1)$th qubit of the right subregister can serve as the target qubit in the required controlled phase. Since the left subregister can store any value $m<k+1$, a separate controlled phase is used for each possibility, where the control bits are given by the values $m$. This leads to a total of $M^3/3 - M^2/2 + M/6$ controlled phase gates for this part of the algorithm. 

By the end of the construction, the phases $A_P$ have been applied to the corresponding permutation label state, having been successively built-up from the elementary transpositions of which they are composed. The explicit circuit for $M=3$ is displayed in Fig.~\ref{fig:permlabelM3circuit}, where now the gates in the red dashed rectangles are included, in order to produce the phases $A_P$. At this point in the construction, the total state of the physical system and the permutation label is
\begin{align}
    \left( \frac{1}{\sqrt{M!}} \sum_P A_P \otimes_{j=M}^1 \ket{Pj}_p \right) \ket{D_{L,M}}_s,
\end{align}
where $\ket{\cdots}_p$ is a state of the permutation label qubits and $\ket{\cdots}_s$ is a state of the system qubits.

In the next step, one applies the position-dependent phase factors $e^{ik_{Pj}x_j}$ to the relevant basis states on the system qubits (step 3 of Algorithm \ref{alg:bethe}).  To do this, we introduce an efficient method that acts on all the $\big( \begin{smallmatrix} L\\ M\end{smallmatrix}\big)$ basis states in $\ket{D_{L,M}}$ simultaneously, yielding an enormous speedup over classical approaches.  The technique, which we call the ``faucet'' method, is based on the observation that the positions $x_j$ take integer values $x_j = 0,\dots,L-1$, so that the total phase $e^{ik_{Pj}x_j}$ can be produced by $x_j$ repetitions of the phase $e^{ik_{Pj}}$.  

For this part of the algorithm, we use the $M$ additional ancilla qubits comprising the faucet register. Each of these new qubits is initialized to $|1\rangle$.  In the outer loop of the faucet subroutine, one traverses the register of the system qubits site by site from $x=0$ to $x=L-1$.  At each site, if it is occupied by a down spin (i.e. the bit is 1), one turns off the next faucet ancilla qubit , $|1\rangle \rightarrow |0\rangle$.  This is achieved through a sequence of multi-controlled $X$ gates, which are controlled on the previous ancilla being in the state $|0\rangle$ and the next one being in the state $|1\rangle$, along with the additional control that the current system site is a $|1\rangle$.  Since the meaning of the ``next faucet ancilla'' at a given site depends on the bitstring, one must generically apply a multi-controlled $X$ gate for each ancilla at every step.  We note that one can decrease the number of gates for sites near the edges of the chain. For instance, at site 2 at most 2 faucets could have been turned off, so that the later ones do not need to be checked. 

\begin{figure}
\includegraphics[scale=0.55]{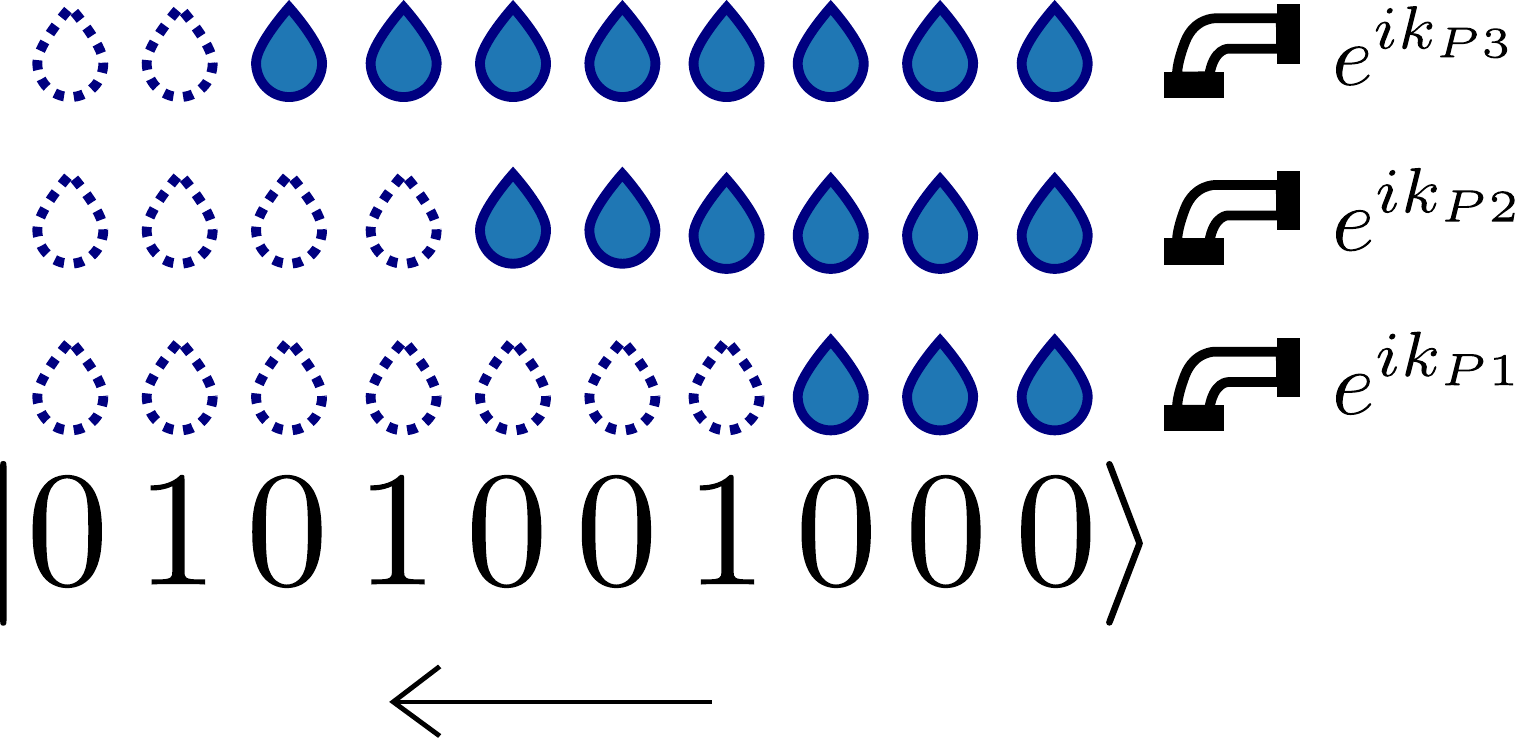}
\caption{ Schematic diagram illustrating the idea of the faucet method. Phases $e^{ik_{Pj}}$ are applied for each system qubit while traversing the bitstring.  When a `1' is encountered, the next faucet in the list is turned off, so that no more phases with the given $k_{Pj}$ value are applied. \label{fig:faucetcartoon}}
\end{figure}

Next, the phases $e^{ik_{Pj}}$ ($j= 1,\dots,M$) are applied to the system qubits, each being controlled on the state of one of the faucet ancillas.  For the $j$th ancilla, this gate is also controlled on the state of the permutation label subregister $j$ (since the value of $k_{Pj}$ is permutation-dependent).  By the end of the system bitstring, all the faucet register qubits are in state $|0\rangle$.  Thus, the subroutine can be compared to a set of $M$ running faucets (that correspond to applying the phases $e^{ik_{Pj}}$), which are turned off at the appropriate times (upon encountering a `1' in the traversal of the system register). This analogy is illustrated in Fig.~\ref{fig:faucetcartoon}. In total, this step requires $M^2L$ doubly-controlled phase gates and a number of Toffoli gates that scales like $\sim M L$ (as mentioned above, some gates can be skipped near the edge of the system).  In our numerical implementation of the method, we introduce additional work qubits to facilitate the construction of these gates using chains of Toffolis \cite{Nielsen2019}.

After the relevant phases have been applied, it remains to disentangle the system from the permutation label (the entanglement having been generated during the faucet method, since the phases there are permutation dependent). This is accomplished by applying the inverse of the circuit that generates the permutation label superposition, without the additional controlled-phase gates that were used to produce the $A_P$ phases.  The fact that the phases $A_P$ and $e^{ik_{Pj}x_j}$ have been applied to the initial Dicke state implies that the permutation label reversal will not completely disentangle the system qubits from the permutation label register.  However, it turns out that the  $|00\dots0\rangle$ component of the permutation state corresponds precisely with the occurrence of the target Bethe ansatz state on the system qubits.  That is, the full state vector takes the form
\begin{align}
|\psi\rangle = \alpha |00\dots 0\rangle_p |\psi_B\rangle_s + \beta |\phi_j\rangle,
\end{align}
where $|\psi_B\rangle_s$ is the normalized target Bethe ansatz state on the system qubits, $|\phi_j\rangle$ is a junk state with $(|00\dots0\rangle\langle00\dots0|_p \otimes \mathds{1}_s)|\phi_j\rangle = 0$.  Thus, by measuring the permutation label qubits, the target Bethe ansatz state is successfully prepared on the outcome  $|00 \dots 0 \rangle$. This result is essentially that obtained from LCU methods~\cite{Berry2015}, though here we have the additional construction of $A_P$ during the label preparation step,which is not present in the standard LCU. To illustrate the full algorithm, the complete circuit to construct a BA eigenstate with $L=4$, $M=2$ is given in Appendix \ref{app:fullcircuitL4M2}. As discussed in our numerical simulations below, the success probability $|\alpha|^2$ depends on the system parameters, and also varies between different eigenstates.  Thus, in general one must repeat the procedure multiple times to obtain the correct state, which can then be used to calculate physical quantities or in other applications, as we discuss in Section \ref{sec:discussion}.  We also show in Section \ref{sec:amplitudeamp} that amplitude amplification can be used to boost the success rate, thereby reducing the overall resource requirements.

\section{Numerical Simulations \label{sec:numerics}}

To calculate the momenta $\{k_i\}$ defining the Bethe eigenstates, we have solved the Bethe equations iteratively using the approach presented in Ref.~\cite{Giamarchi2004}. We then performed numerical simulations of Algorithm \ref{alg:bethe} using the IBM Qiskit library's state vector simulator \cite{Qiskit2019}.  These calculations verify the correctness of our algorithm, and reveal its success probabilities for the sufficiently small systems that can be studied on a classical computer.  However, we can also explicitly construct the circuits that would need to be run for much larger instances, far beyond what is classically tractable.  The corresponding circuit depths and gate counts in these cases indicate that our algorithm is feasible for near-term error-corrected quantum computers.  We stress that our analysis does not rely on asymptotic resource scaling arguments, but rather provides exact gate counts, since the corresponding circuits are precisely known.  Although we have not compiled our algorithm down to an error-correcting code such as the surface code, we estimate the required number of T gates below.

\begin{figure}
\includegraphics[scale=0.58]{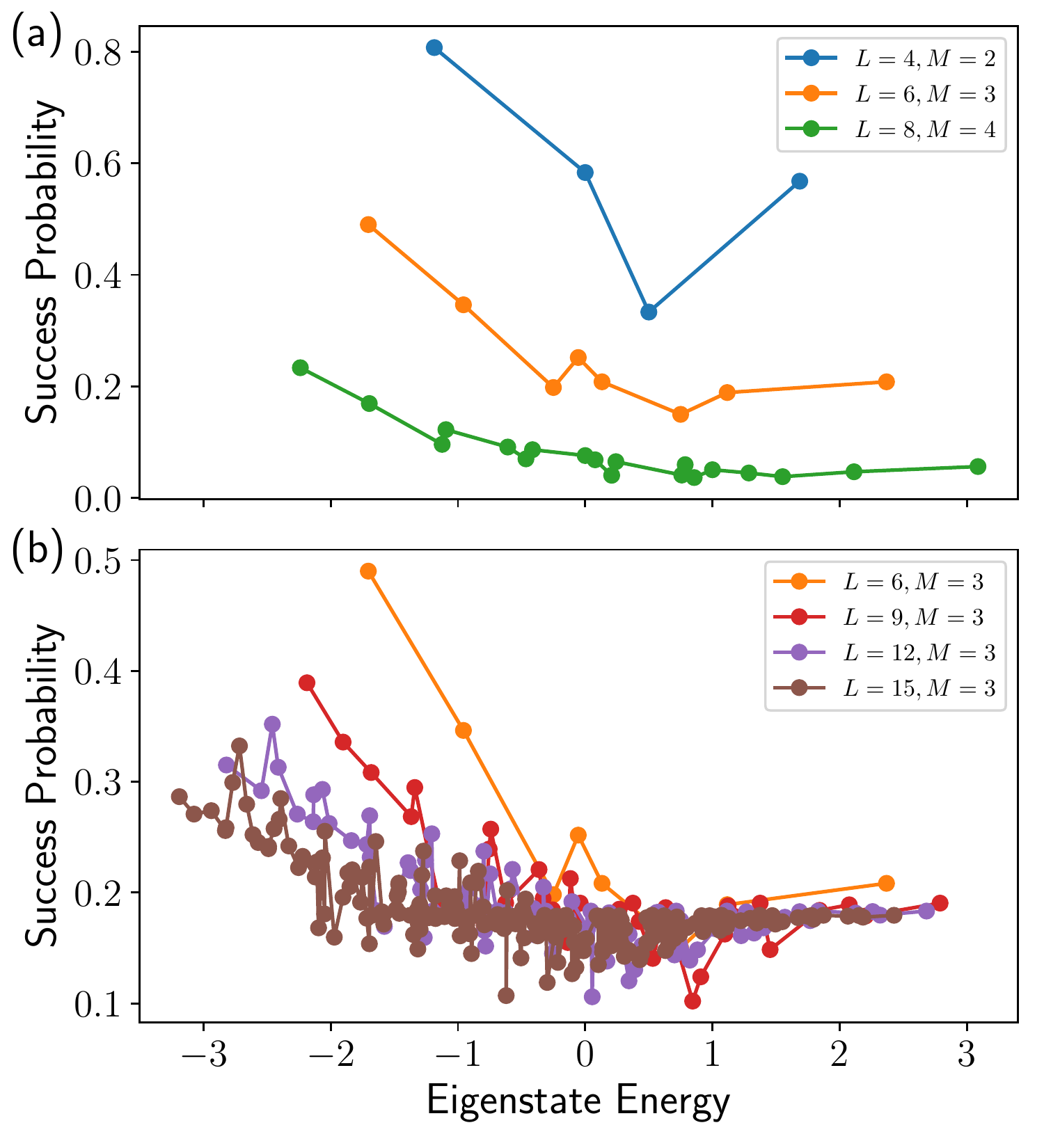}
\caption{ (a) Success probabilities for preparation of selected eigenstates as a function of their energies, when $L=2M$ and $M=2,3,4$. (b) Success probabilities of selected eigenstates for $M=3$ and varying $L$. Hamiltonian parameter values are $J_{xy}=1$ and $J_z = -1/2$.\label{fig:algo3successprob}}
\end{figure}

In Fig.~\ref{fig:algo3successprob}(a) we present the numerically-calculated success probability of the algorithm for preparing selected eigenstates when $L=2M$ and $M=2,3,4$, with $J_{xy}=1$, $J_z=-1/2$. The interaction strength in this case corresponds to the critical regime of the ferromagnetic model. We note the eigenstates included in Fig.~\ref{fig:algo3successprob} are not meant to comprise the complete set of real-valued solutions, but rather are simply those for which we obtained numerical solutions of the Bethe equations by using the algorithm presented in Ref. \cite{Giamarchi2004}. Two general trends are apparent: a significant suppression of the success rate with increasing $M$, and a more moderate suppression as a function of the eigenstate energy, within each set of system parameters $L$, $M$.  The worst-case probabilities are roughly consistent with $1/M!$, although we have only limited values of $M$ to support this (larger $M$ being outside of our computational resources for classical simulation).  This value is further supported by Fig.~\ref{fig:algo3successprob}(b), which shows the success probabilities for various eigenstates when $M=3$ and for different values of $L$, with the clear trend that increasing $L$ tends to flatten the success probability across the spectrum. Although the low energy states enjoy less of an advantage over the higher energy ones in this case, the lowest probabilities are still around $1/M!$ on average. However, the behavior of the success rate changes significantly depending on the value of $J_z$. These effects are considered Appendix \ref{app:varyJz}.

These results suggest one can go to very large system sizes $L$, while still preserving a reasonably large success probability, if $M$ is sufficiently small.  We note that the case of small $M$ is particularly interesting for physics applications in the ferromagnetic regime of the model.  In this case, the all-up state $\ket{0}^{\otimes L}$ is a ground state of the model, while small $M$ states are low-lying excited states of interacting magnons (spin waves). The ability to study these states as a function of $M$, as enabled by our algorithm, would yield deeper insight into the development of strong correlations in these systems as the number of interacting quasiparticles grows. While the ferromagnetic regime is especially natural for our algorithm, we note that interesting physics in the paramagnetic and antiferromagnetic regimes can also be explored at small values of $M$.  These correspond to highly-excited eigenstates, which are relevant, for instance, in the study of quantum thermalization \cite{Essler2016,Vidmar2016}.

\begin{figure*}
\includegraphics[scale=0.75]{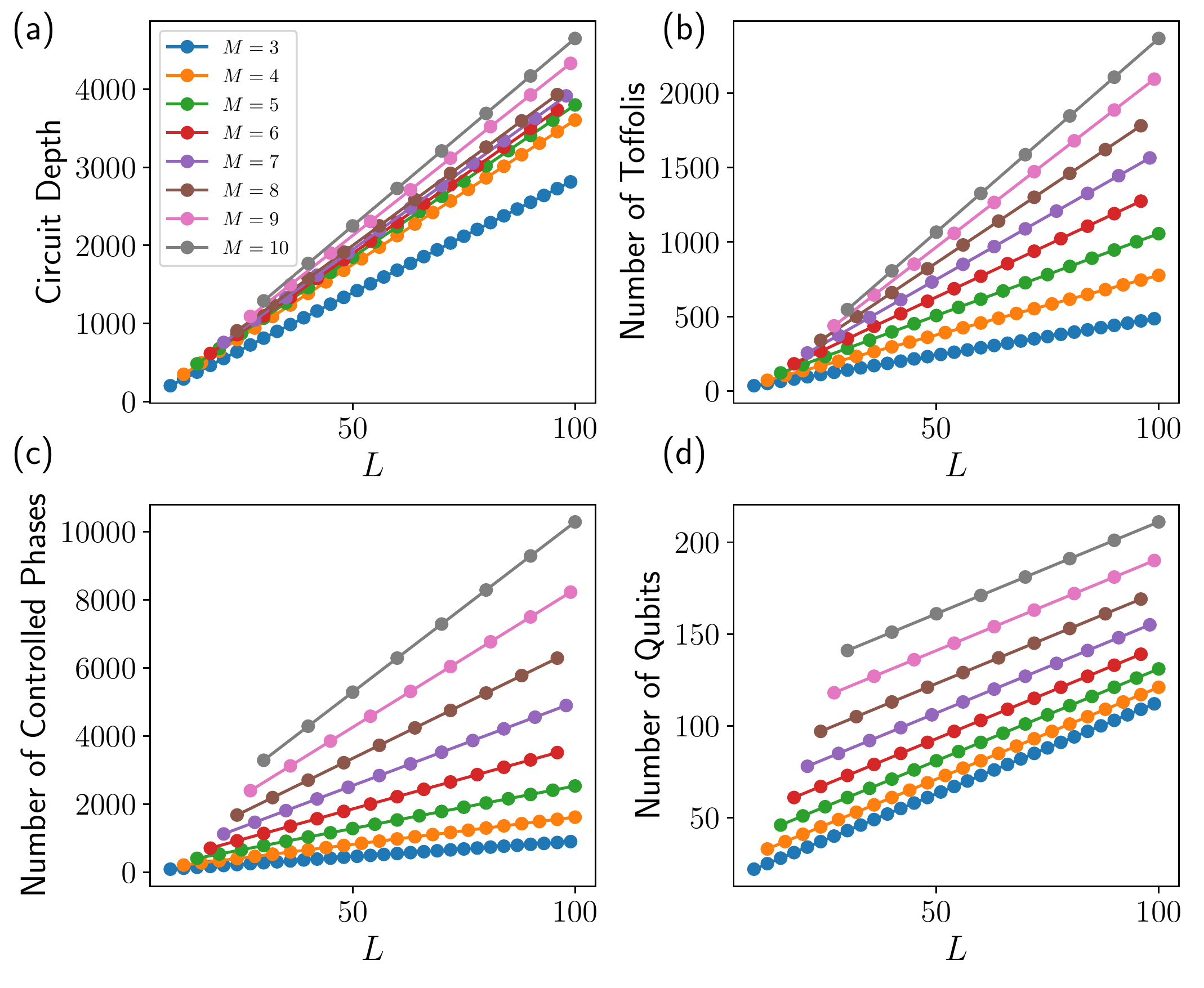}
\caption{ Bethe state preparation (a) circuit depth, (b) number of Toffoli gates,  (c) number of controlled phase gates, and (d) number of qubits versus system size $L$, for different numbers of down spins $M$. \label{fig:algo3gatecounts}}
\end{figure*}

For these applications (and others discussed below), it appears feasible to access values of $L$ and $M$ that would not be classically simulable (even by approximate methods), while maintaining relatively modest resource requirements for the algorithm.  This is demonstrated in Fig.~\ref{fig:algo3gatecounts}, which provides circuit depths [Fig.~\ref{fig:algo3gatecounts}(a)], Toffoli gate counts  [Fig.~\ref{fig:algo3gatecounts}(b)], controlled phase gate counts  [Fig.~\ref{fig:algo3gatecounts}(c)], and the number of qubits required  [Fig.~\ref{fig:algo3gatecounts}(d)] for the Bethe state preparation algorithm.  The linear scaling of all these metrics in $L$ is immediately apparent.  The slopes of these lines for the case $M=5$ are (a) 39, (b) 11, (c) 25, and (d) 1, respectively.  The results for the number of controlled-phases and qubits are in exact agreement with the analytical results of Section \ref{sec:bethealg}.

\begin{figure}
\includegraphics[scale=0.43]{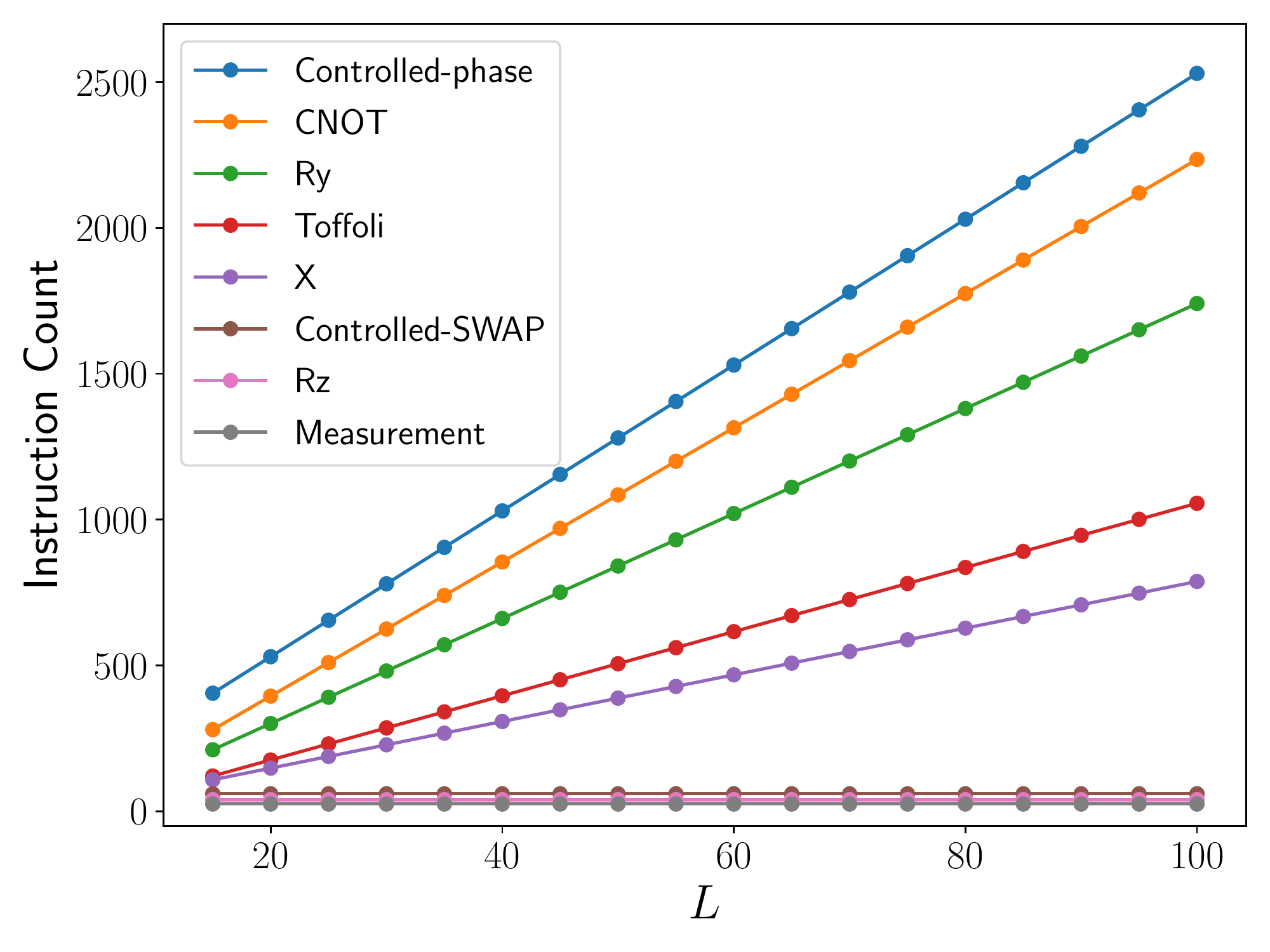}
\caption{ Bethe state preparation gate and measurement counts for $M=5$ as a function of $L$. \label{fig:algo3totalgatecountsM5}}
\end{figure}

Apart from the asymptotic scaling behavior, the absolute values of the circuit depths and gate counts are seen to be very low, on the order of $10^3$--$10^4$, even for large systems of $L\sim100$ sites.  Furthermore, the total number of qubits required ($\sim 10^2$) is also quite reasonable for small error-corrected devices.  In Fig.~\ref{fig:algo3totalgatecountsM5} we show the total gate and measurement counts for the case $M=5$ as $L$ is varied.  This indicates that the controlled-phase, $R_y$, and Toffoli gates are the most prevalent non-Clifford operations in the algorithm.  To estimate the fault-tolerant resources needed, we therefore convert the counts for these gates into the corresponding numbers of T gates.  Following Ref. \cite{Reiher2017}, we assume a worst-case scenario for the number of T gates needed to realize an arbitrary $z$ rotation to be given by $4\log_2 (1/\epsilon)+11$, where $\epsilon$ is the rotation synthesis error \cite{Selinger2015}.  Similarly, arbitrary $y$ rotations can be performed by conjugation with Clifford operations. As in Ref. \cite{Kivlichan2020}, we replace each Toffoli gate with two T gates \cite{Gidney2019}.  For $L=100$, $M=5$ this leads to a T count of $\sim 6.2 \times 10^5$ for a single run of the state preparation algorithm, with $\epsilon=10^{-10}$.  Assuming a worst-case success probability of $1/M!$, approximately 120 attempts would need to be performed on average to correctly generate the target eigenstate.  This yields $\sim 7.4 \times 10^7$ T gates overall, which is comparable to the estimates for simulating the Hubbard model using the methods of Ref. \cite{Kivlichan2020}.  We note that the estimates in that work involve optimizing an error budget between multiple sources (Trotterization, phase estimation, and rotation synthesis), and do not appear to include the cost of preparing a good initial state for the phase estimation routine.  Furthermore, the above estimate for our algorithm assumes the seemingly worst-case scenario in the number of repetitions ($\sim M!$), whereas the results in  Fig.~\ref{fig:algo3successprob} suggest that lower energy states require fewer repetitions in general. To reduce the number of repetitions required for our algorithm, we implement amplitude amplification in the following section.

\section{Amplitude Amplification \label{sec:amplitudeamp}}

\begin{figure}
\includegraphics[scale=0.55]{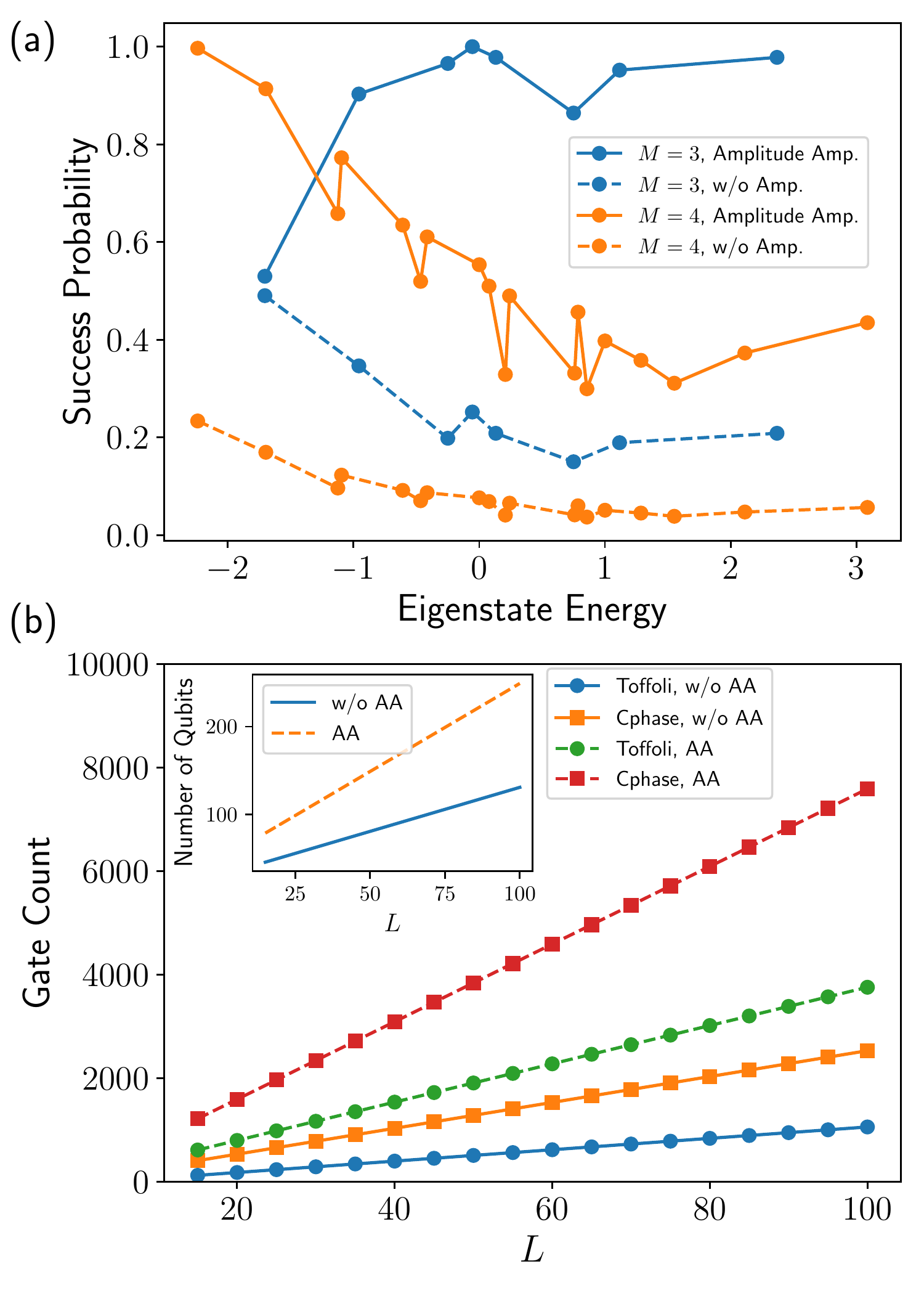}
\caption{(a) Success probability for Bethe state preparation with (solid lines) and without (dashed lines) amplitude amplification.  In the former case, a single round of amplification is applied.  System parameters are $J_{xy}=1$, $J_z = -1/2$, $L=2M$. (b) Gate counts as a function of system size $L$ for Toffoli (circles) and controlled-phase (squares) gates without (solid lines) and with one round of amplitude amplification (dashed lines). (Inset: total number of qubits without (solid) and with (dashed) amplitude amplification).\label{fig:amplitudeamp}}
\end{figure}

Amplitude amplification, a generalization of the well-known Grover search algorithm, is a quantum subroutine which can boost the probability of a desired measurement outcome, leading in general to a square root improvement in the number of repetitions required for the success of a probabilistic algorithm \cite{Brassard2000}.  For our problem, we use $\mathcal{B}$ to denote Algorithm \ref{alg:bethe} with the measurement step removed.  Amplitude amplification defines an operator 
\begin{align}
\mathcal{Q} = - \mathcal{B} S_0 \mathcal{B}^{-1} S_B,
\label{eq:ampamp}
\end{align}
where $S_B$ changes the relative sign of the ``good'' states in the Hilbert space, while $S_0$ changes the relative sign of the vacuum state $|00\dots0\rangle$.  In the present case, the good states are the components of the Bethe ansatz state, which correspond to $|00\dots0\rangle_p$ on the permutation label qubits.  $S_B$ can therefore be implemented using a \textsc{OR} circuit on the permutation label, followed by $Z$ on the work qubit that stores the result, after which the \textsc{OR} is uncomputed.  In our numerical calculations of the success probability, we use the ancilla-free implementation of \textsc{OR} in Qiskit's standard circuit library.  The ancilla-free approach is used here to decrease the number of qubits needed for the simulation, allowing us to study larger system sizes. Below we examine an ancilla-based method for which the gate counts at large sizes are reduced. To produce $S_0$ we use the same approach, with the \textsc{OR} circuit extended to include the system qubits (we do not implement the $-1$ in Eq. \eqref{eq:ampamp}, as it is an overall phase).  We present numerical results for amplitude amplification in Fig.~\ref{fig:amplitudeamp}(a).  This confirms the clear enhancement of the algorithm success probability using this method.  Although only one round of amplification has been applied here, the protocol can be repeated in the standard way to further increase the success rate.  Applying this improvement to the resource estimate of the previous section, the $M=5$ worst-case eigenstates should require on average $\sqrt{120}\approx 11$ repetitions of the algorithm $\mathcal{B}$ to achieve success, leading to an overall T count of $\sim 4.1\times 10^6$ (neglecting the costs of $S_B$ and $S_0$). 

To estimate the resources needed for amplitude amplification more precisely, we implement the multi-controlled \textsc{OR} operation using elementary gates and ancillas \cite{Nielsen2019}. The Toffoli count, controlled-phase count, and number of qubits are shown in Fig.~\ref{fig:amplitudeamp}(b) for a single round of amplification when $M=5$. Since the $S_0$ reflection depends on the state of both the system and the permutation label, the implementation requires an additional $L + M^2 - M$ ancillas (we can re-use the $M$ faucet ancillas for the amplification). The additional gates required are dominated by the cost of executing $\mathcal{B}$ three times. Although the total number of qubits is approximately doubled in this approach, we note that it may be possible to reduce this number by acting with the \textsc{OR} operation on only a subset of the qubits that are nominally necessary to identify the state. For example, although $S_B$ is an \textsc{OR} circuit acting on the $M^2$ permutation label qubits, in practice the computational basis states appearing in the junk state $|\phi_j\rangle$ can be distinguished from $|00\dots0\rangle_p$ by only acting on a reduced number of qubits. While the particular subset of qubits needed will depend on the state under consideration, it is possible to confirm the success of this approach by measuring the energy or other quantities that can be compared to exact analytic expressions.

We have also implemented a different version of amplitude amplification, which is a modified form of the oblivious amplitude amplification of Ref.~\cite{Berry2015}.  Unfortunately, this method leads to a reduced fidelity of the actually prepared state with the exact target state, though in some cases the overlap remains quite high ($>0.99$).  We attribute this reduced fidelity to the non-unitarity of summing exponentials with unit modulus, since $|e^{ia} + e^{ib}| \neq 1$ in general.  We note, however, that nearly deterministic success of the oblivious amplitude amplification procedure was obtained for the application of Ref.~\cite{Berry2015} (simulation of Hamiltonian dynamics with Taylor series expansions).  For this reason, it is less clear how the approach will fare for the Bethe state preparation problem at larger values of $M$.  In addition, further modification to the method of Ref.~\cite{Berry2015} may ameliorate some of the difficulties with applying it to Bethe state preparation in its present form.

\section{Comparison with Alternative algorithms \label{sec:comparealg}}

To highlight the advantages of Algorithm \ref{alg:bethe}, we compare it with conceptually simpler, but much less efficient, methods of preparing Bethe ansatz states on a quantum computer.  First, one could imagine applying controlled phase gates directly to each term in the Dicke state superposition to generate the desired eigenstate.  This approach still requires permutation label ancillas to generate the linear combination of phases needed in Eq.~\eqref{eq:bethewavefunction}.  However, it has the seeming advantage of allowing one to combine the phases $A_P$ and $e^{ik_{Pj}x_j}$ into a single controlled-phase rotation, whereas the former term required order $M^3$ and the latter one order $M^2$ controlled-phases to implement using Algorithm \ref{alg:bethe}.  Nevertheless, it is clear that this benefit is vastly outweighed by the large number of terms in the superposition that must be separately addressed, $M! \big( \begin{smallmatrix} L\\ M\end{smallmatrix}\big)$.  For the case $L=100$, $M=5$ this amounts to $\sim 9.0 \times 10^9$ controlled phases, compared to the $2530$ of Algorithm \ref{alg:bethe}.

\begin{figure}
\includegraphics[scale=0.52]{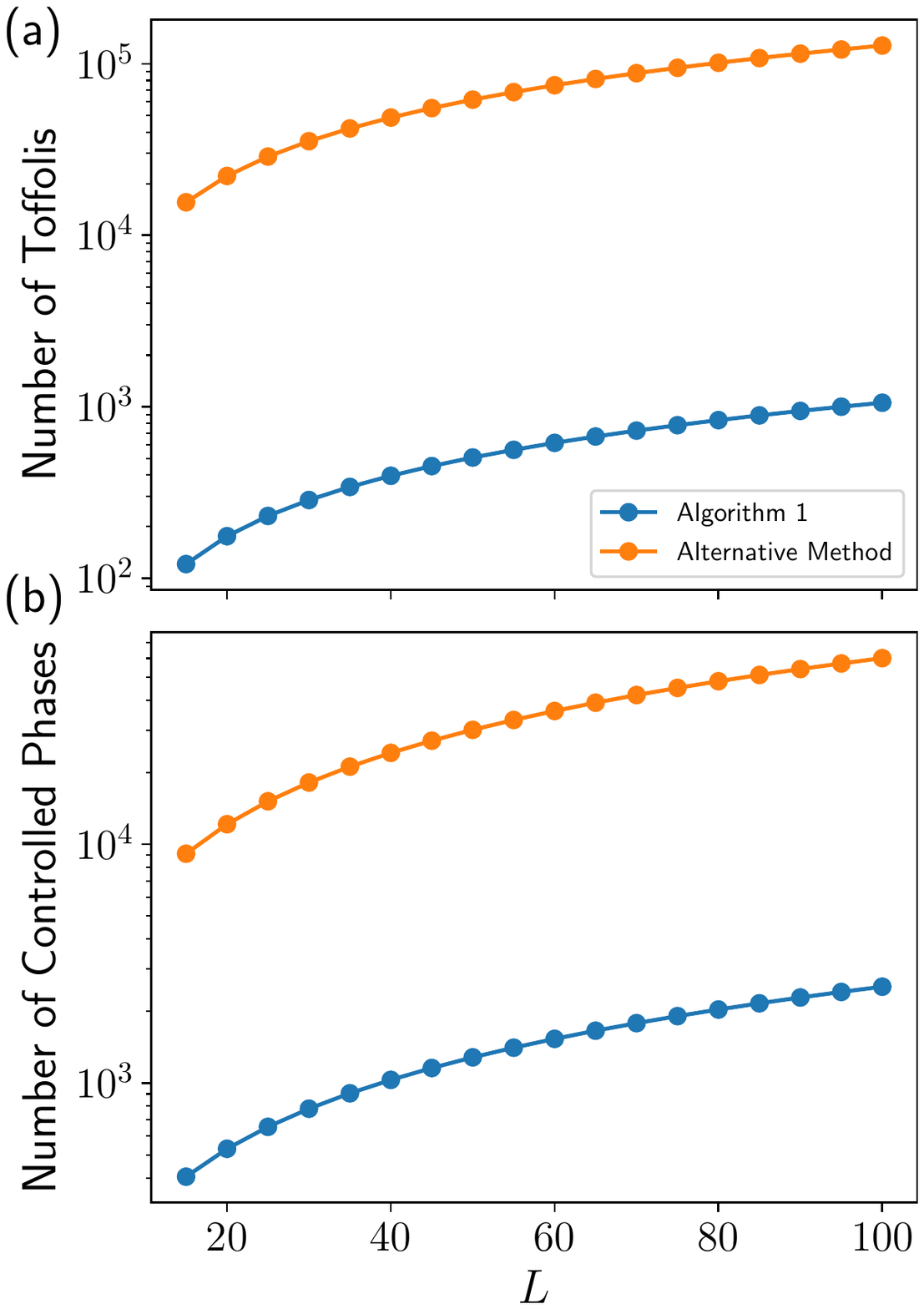}
\caption{ Comparison between Algorithm \ref{alg:bethe} and the alternative method of the number of (a) Toffoli and (b) controlled-phase gates used, for $M=5$. \label{fig:algo2vs3}}
\end{figure}

A more promising approach is to use the ``faucet'' method of Algorithm \ref{alg:bethe} to handle the $e^{ik_{Pj}x_j}$ phases while still applying the full $A_P$ in a single multi-controlled-phase gate, rather than decomposing it into its elementary transpositions.  This replaces the $M! \big( \begin{smallmatrix} L\\ M\end{smallmatrix}\big)$ dependence above with $M! L M$.  While the scaling is still inferior to that of Algorithm \ref{alg:bethe} for large $M$, it is conceivable that for small $M$ this alternative method may be competitive.  In particular, one can replace the complicated permutation label of Algorithm \ref{alg:bethe}, which required $M^2$ qubits to construct $A_P$ in terms of individual transpositions, with a compressed label that simply assigns a number to each permutation.  This approach uses significantly fewer qubits, at the expense of requiring more controls for the relevant phase gates.  Since the permutation label construction still needs to be reversed to disentangle the system from the ancillas, it is important that it can still be executed in a unitary fashion.  This in turn requires an efficient method for generating an equal superposition of $M!$ states.  We implement such states using the  prime factorization $M! = 2^{n_2} 3^{n_3} \cdots$.  We then construct the permutation label as the tensor product of the binary representation of $2^{n_2}$ (using $n_2$ qubits) and $W_n$ states for the odd prime factors.  The equal superposition for the binary part of the label is easily generated by applying H gates to the relevant qubits, while various efficient algorithms exist for $W_n$ state preparation \cite{Yesilyurt2016,Cruz2019,Gidneystackexchange2018}.  We implemented this algorithm numerically and verified that it successfully prepares Bethe ansatz eigenstates.  Since the fundamental approach for creating the linear combination of phases is the same between this method and Algorithm \ref{alg:bethe}, their success probabilities are equal.  Unfortunately, explicit construction of the circuits for the alternative method indicates that the resource requirements are significantly higher, even for small $M$.  This is shown in Fig. \ref{fig:algo2vs3}, which reveals that the number of controlled phase and Toffoli gates required for the alternative method vastly exceeds those of Algorithm \ref{alg:bethe}, even for $M=5$.

It is also useful to compare Algorithm \ref{alg:bethe} with other more standard techniques, such as adiabatic state preparation \cite{Farhi2000,Reichardt2004,Aspuru-Guzik2005}. As is well-known, the evolution time to produce a large overlap with the desired eigenstate is expected to scale as the inverse square of the gap between the energy of the given state and that of the next closest one. This time can be short, for instance, in the ordered regime of the XXZ chain, where the gap between the ground and first excited state remains finite in the large system limit.  However, in the critical regime of the model the spectrum is gapless, and so the evolution time is expected to diverge.  Furthermore, even in the ordered regime the spectrum at higher energies above the ground state generically has dense regions, preventing an efficient preparation of those eigenstates by the adiabatic approach.  Our algorithm does not suffer from such complications, as it implements the exact analytic expression for the wave function, irrespective of the gap between the target eigenstate and the other ones.

\section{Discussion \label{sec:discussion}}

To achieve quantum advantage for a physically relevant problem, it should be the case that no classical method could deliver results of a comparable accuracy.  Since the present quantum algorithm exactly prepares eigenstates of the XXZ chain, it is reasonable to compare it to the numerical exact diagonalization of finite-size systems (i.e. using classical computers).  In a recent study, a matrix-free approach was used to investigate Heisenberg spin chains up to length $L=26$ \cite{VanBeeumen2020}.  This work had vastly reduced the memory requirements compared to conventional methods, though the scaling remained exponential with system size.  Specifically the $S^z = 0$ subspace ($M=13$) was considered, for which the dimension is $\sim 10^7$.  In contrast, the $L=100$, $M=5$ subspace is roughly seven times larger (dimension $\sim 7.5\times10^7$).  Although state vectors of this size can still be stored in memory, we note that the computation time is also exponential in the system size, ultimately limiting the practicality of exact diagonalization.

In addition to numerically exact calculations, approximate tensor network methods have been highly successful for studying one-dimensional quantum many-body systems with the matrix product state (MPS) ansatz \cite{Verstraete2006,Verstraete2008}.  However, these approaches are best-suited for states with a relatively low amount of entanglement, such as gapped ground states obeying an area law for the entanglement entropy.  This makes simulation of long-time dynamics challenging, due to the growth of entanglement from, for instance, an initial product state.  Our Bethe ansatz algorithm can prepare eigenstates throughout the spectrum at the same computational cost, including highly excited states whose entanglement entropy grows more quickly than logarithmically, even in the small $M$ limit \cite{Molter2014}.  This yields an advantage over MPS methods for large systems when targeting these strongly entangled states. An explicit link between the Bethe ansatz and MPS was developed in Refs. \cite{Murg2012,Chong2015}, which used the algebraic Bethe ansatz to produce exact tensor network representations for generic eigenstates. These networks have a PEPS-like structure (but with fewer physical indices), which underscores the computational intractability of such states for large systems.

In addition to computing arbitrary-range and higher-order correlation functions that are inaccessible with traditional Bethe ansatz methods, our algorithm has a number of other applications.  Simulation of the real-time dynamics of many-body systems is widely recognized as a task allowing for quantum advantage.  Such simulations often take the form of quench experiments, for which the system is initialized in an easy-to-prepare product state, then allowed to evolve under the influence of an interacting many-body Hamiltonian.  Our algorithm would enable interesting variations on this approach, for instance by initializing the system in an eigenstate of a given value of the interaction strength, then subsequently evolving it with a different value.  The evolution here can be performed using any of the known algorithms for quantum simulation, whether by Trotterization \cite{Whitfield2011,Hastings2015,Kivlichan2020}, Taylor expansions \cite{Berry2015,Babbush2016}, or other approaches \cite{Childs2012,Low2019}.  

In a different direction, one could use our algorithm as a starting point to explore integrability-breaking perturbations.  Thus, we consider a Hamiltonian of the form $\mathcal{H} = \mathcal{H}_0 + \mathcal{H}_p$, where $\mathcal{H}_0$ is solvable by the Bethe ansatz and $\mathcal{H}_p$ includes perturbations that break the integrability of the total Hamiltonian $\mathcal{H}$ (such as disorder in the coupling strengths).  In this case, the Bethe state prepration algorithm is used to prepare an eigenstate of $\mathcal{H}_0$, which then serves as an initial state for quantum annealing or phase estimation on $\mathcal{H}$.  For sufficiently weak perturbations, the overlap of this state with the corresponding exact eigenstate of $\mathcal{H}$ should be much greater than that of a mean-field or non-interacting trial state.

Although we have focused on deploying our algorithm on small error-corrected quantum computers, one may also consider implementing it on present-day or near-term NISQ devices.  Many of the controlled-phase gates in our algorithm involve repetitions of the same basic rotation angles, of which there are only $(M^2 + M)/2$ distinct values.  This suggests replacing the exact values with variational parameters, similarly to Ref. \cite{Nepomechie2021}.  The resulting variational form can then be optimized under the cost function $|E-E_B|$, where $E$ is the energy calculated on the quantum computer and $E_B$ is the exact value, known analytically from the Bethe ansatz solution.  We note that the present optimization problem should be significantly easier than that of a standard VQE, since the ideal values of the rotation angles can serve as a good initial guess.  Updates to the parameters then serve to directly mitigate systematic errors due to over- or under-rotation in the controlled-phase gates.

\section{Conclusions \label{sec:conclusions}}

We have presented a quantum algorithm for the efficient preparation of Bethe ansatz eigenstates of the XXZ model.  To our knowledge, this is the first quantum algorithm for the direct preparation of eigenstates of an interacting many-body problem.  The circuit depth and gate counts of the algorithm scale linearly in the system size, for a fixed number of down spins.  Our algorithm is feasible to perform on small error-corrected devices of order 100 qubits, provided that the number of down spins is small.  The usefulness of the approach can be extended through amplitude amplification.  In particular, quantum advantage over classical computational methods appears to be achievable, with resource estimates that are comparable to the most efficient known quantum simulation algorithms.  Our work suggests directions for future research, including the modification of the algorithm for the case of complex-valued $\{k_i\}$, and its generalization to other Bethe ansatz-solvable models, such as the one-dimensional Hubbard model.

\begin{acknowledgments}
Numerical simulations were performed with IBM Qiskit and the QuSpin Python library \cite{Weinberg2017}. We would like to thank Chandra Sekhar Mukherjee, Jesko Sirker, and Rafael Nepomechie for helpful discussions. This work was supported by the Department of Energy. E.B. and N.J.M. acknowledge Award No. DE-SC0019199, and S.E.E. acknowledges the DOE Office of Science, National Quantum Information Science Research Centers, Co-design Center for Quantum Advantage (C2QA), contract number DE-SC0012704.
\end{acknowledgments}

\appendix
\section{Full circuit for $L=4$, $M=2$ \label{app:fullcircuitL4M2}}
\begin{figure*}
\includegraphics[scale=0.35]{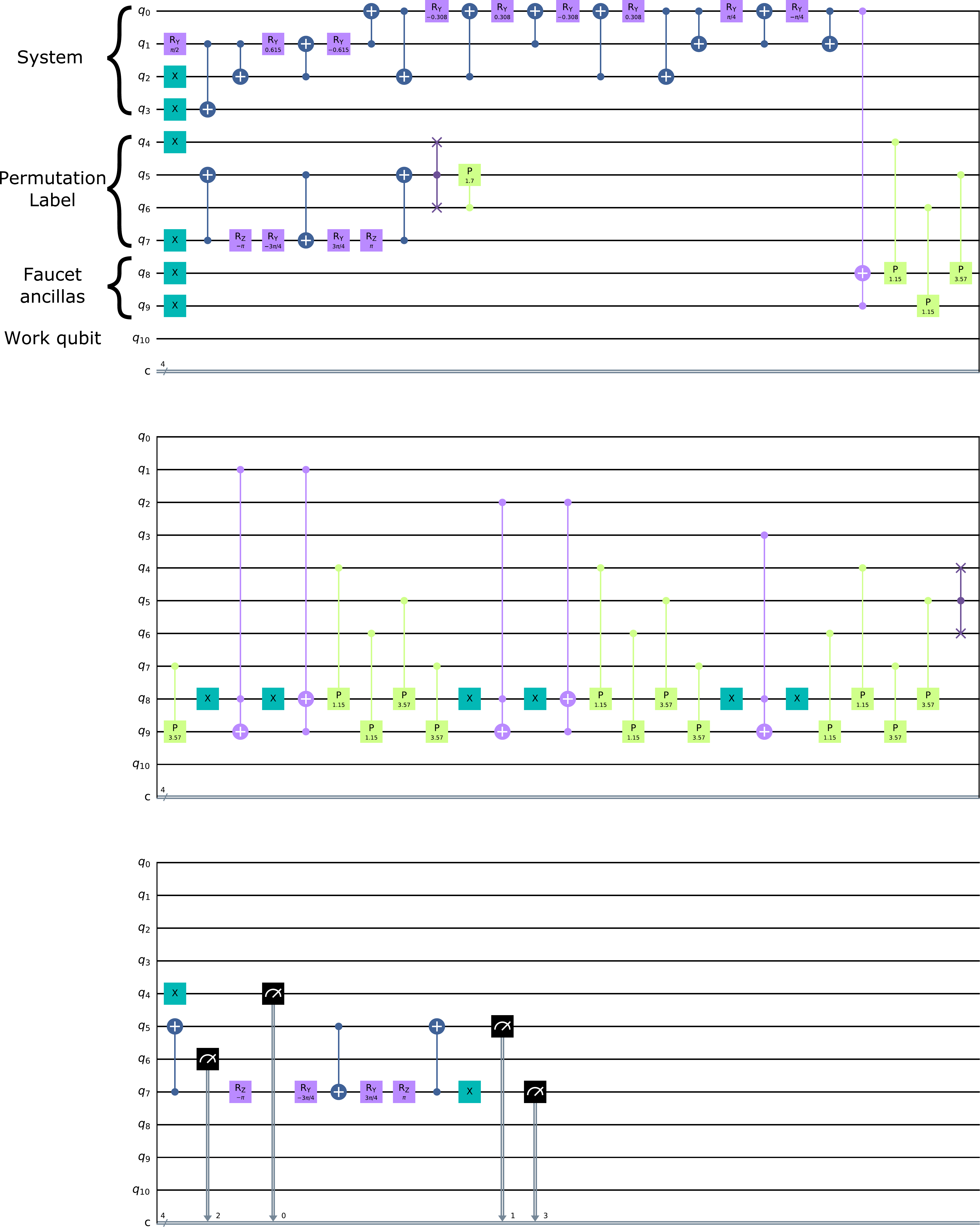}%
\caption{Quantum circuit to prepare the Bethe ansatz eigenstate with $L=4$, $M=2$, $k_1 = 1.14676529$, $k_2 = 3.56562369$.\label{fig:fullcircuitL4M2}}
\end{figure*}

In Fig.~\ref{fig:fullcircuitL4M2} we present the full quantum circuit to prepare a Bethe ansatz eigenstate with $L=4$ and $M=2$. We note that the work qubit is not needed when $M=2$ case, but we have included it here as a reminder that it is used to implement multi-controlled gates for $M>2$.

\section{Success Probability for different $J_Z$ \label{app:varyJz}}
\begin{figure}
\includegraphics[scale=0.54]{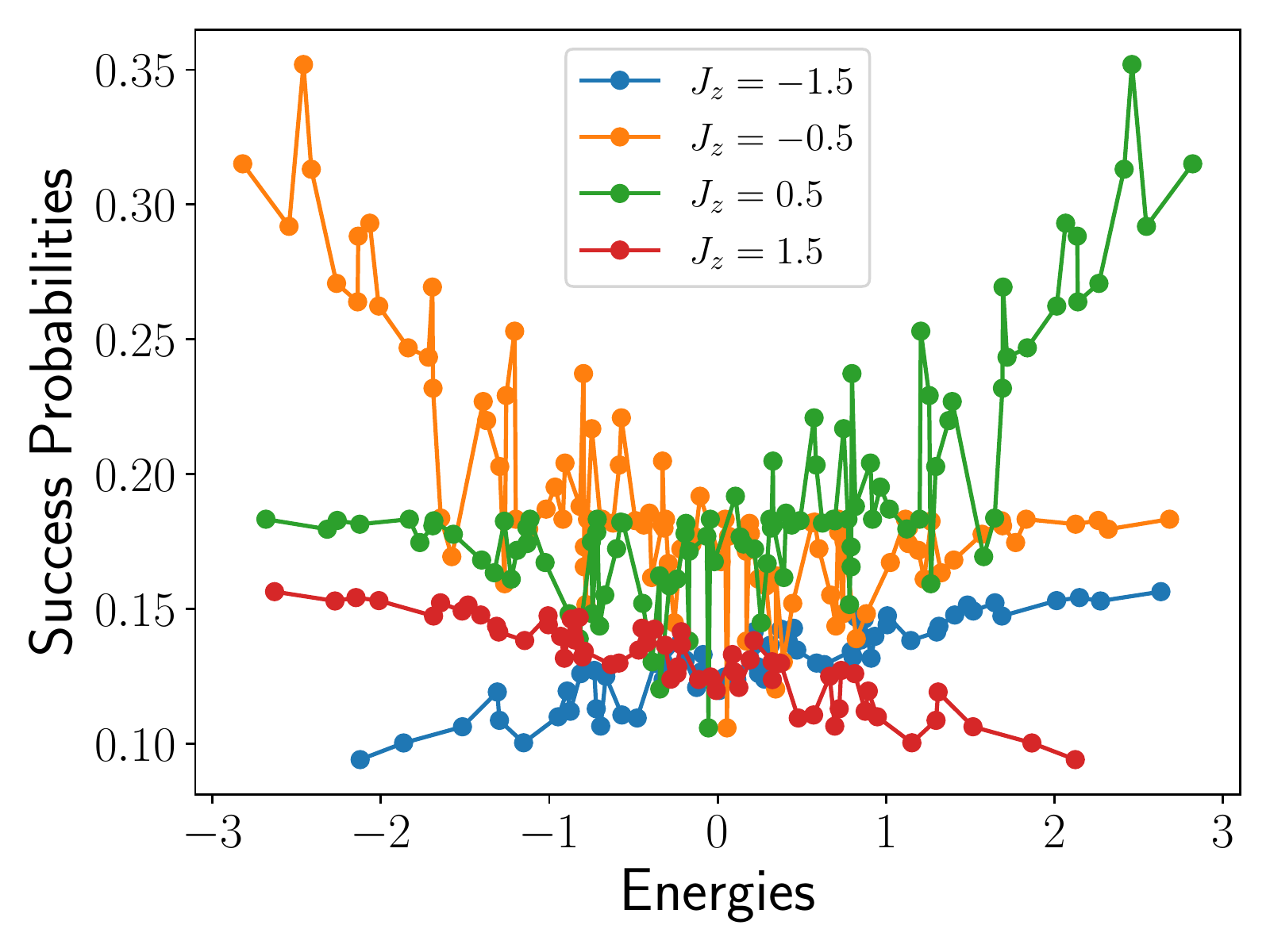}
\caption{Success probabilities for preparation of selected eigenstates as a function of their energies, for $L=12$, $M=3$, $J_{xy}=1$, with varying interaction strengths $J_z$. \label{fig:varyJz}}
\end{figure}

The success rate of the Bethe ansatz state preparation changes qualitatively with both the sign of $J_z$ and whether it lies in the critical ($|J_z| < 1$) or non-critical ($|J_z| \geq 1$) regimes. These effects are illustrated in Fig.~\ref{fig:varyJz}. Whereas the ferromagnetic model in the critical regime has higher success probabilities for lower energy states, the antiferromagnetic case shows the opposite behavior.  In fact, the success probabilities for different states are exactly mirrored across the $E=0$ axis. This is especially interesting since the corresponding eigenstates at $E=-E_i$ for the FM case and $E=E_i$ for the AFM one are different in general.  It is unclear at present why these distinct states should have the same success probability. When $J_z$ is outside the critical regime, we find that the success probability can be lower than the $1/M!$ value ($\approx 0.17$ for $M=3$) that appears to bound the critical case.

% uncomment for PRX Quantum formatting
\bibliography{bibbetheansatzstateprep}

% uncomment for Quantum journal fomatting
% \bibliographystyle{unsrtnat}
% \bibliography{bibbetheansatzstateprepoverleaf}
\end{document}